# Floquet states on disclinations

K. Sabour,[1,*] S. K. Ivanov,[2,*] A. Ferrando,[2] and Y. V. Kartashov[3]

[1]*Moscow Institute of Physics and Technology, Institutsky lane 9, Dolgoprudny, Moscow region, 141701, Russia*
[2]*Instituto de Ciencia de los Materiales, Universidad de Valencia, Catedrático J. Beltrán, 2, 46980, Paterna, Spain*
[3]*Institute of Spectroscopy, Russian Academy of Sciences, 108840, Troitsk, Moscow, Russia*
*Corresponding authors: Khalilsabor1998@gmail.com, Sergei.Ivanov@uv.es

We show that periodic longitudinal modulation of a waveguide arrays with disclination can result in the appearance of previously unexplored Floquet modes bound to the disclination core. Such modes arise due to oscillations of the waveguides in the array, periodically switching the structure between topological and trivial phases on each modulation period, so that on average it seems trivial. Localization of such modes depends on the amplitude of waveguide oscillations. Depending on the discrete rotational symmetry of the arrays with disclinations, these modes exhibit distinct spatial profiles unattainable in periodic lattices. Propagation in a medium with focusing cubic nonlinearity reveals that these Floquet states remain localized below a critical power threshold, indicating on the possibility of formation of disclination-bound Floquet solitons. Our results unveil a new regime of localization in photonic systems, bridging disclination topology, Floquet engineering, and nonlinearity.

Edge states with tunable localization properties and propagation velocities arising in materials with nontrivial band topology illustrate unique opportunities offered by such materials for controlling edge currents, their routing and protection. Following their initial discovery in electronic systems, such materials have been realized across diverse physical platforms, including photonic structures [1,2]. Conventional $d$-dimensional topological insulators support at their boundaries $(d-1)$-dimensional in-gap edge states, localized in the direction transverse to the boundary, whose number is directly linked with topological indices of the bands. More recently, a new class of higher-order topological insulators (HOTIs) has been proposed [3,4]. Their distinctive feature is that an $n$-th order HOTI may host boundary states (including surface, edge, or corner ones) of reduced dimensionality down to $(d-n)$. In these structures, that are usually periodic in their bulk, topological phases arise from tailored shifts of sites within the unit cell that modify intra- and intercell couplings.

Recently, it was shown that higher-order topological phases can also arise in aperiodic structures containing disclinations – defects that globally modify the lattice structure and produce systems with distinct discrete rotational symmetries and properties unattainable in periodic HOTIs [5,6]. Such defects can trap fractional spectral charges, a property manifested in the appearance of localized modes at the disclination core of the structure [7,8], in addition to conventional edge and corner states that may arise at its outer edges [9,10]. The concept of HOTIs with disclinations has been further extended to non-Hermitian lattices [11]. Photonic systems can exhibit strong nonlinear response which, when combined with the nontrivial topology of structures with disclinations, enable the formation of nonlinear disclination states and vortices in conservative settings [12, 13]. For realization of linear vortex disclination states see [14].

Beyond static implementations, periodic modulation of system parameters in evolution variable can profoundly reshape the modal spectrum [15] and even fundamentally alter topological properties of the system giving rise to Floquet topological phases [1,2]. Remarkably, periodic modulation can also lead to the emergence of anomalous Floquet $\pi$ modes, arising only due to modulation, whose localization strongly depends on the modulation parameters [16-18]. Such states have been experimentally demonstrated in conservative [19] and dissipative [20] photonic Su-Schrieffer-Heeger lattices with dynamically modulated intra- and inter-cell couplings. The inclusion of nonlinearity in such Floquet systems enables the formation of topological $\pi$-solitons – unique states [21] combining topological protection with nonlinear self-action, that were studied very recently, both theoretically [22, 23] and experimentally [24]. Furthermore, envelope solitons have been studied in Floquet valley-Hall photonic lattices [25]. Stable lasing in $\pi$ modes emerging in one-dimensional non-Hermitian Su-Schrieffer-Heeger arrays has also been predicted in [26].

These two classes of systems – modulated Floquet structures and aperiodic arrays with disclinations – enable the exploration of intriguing topological phenomena but have so far been studied independently owing to the fundamentally different mechanisms underlying edge state formation in each system. While non-Hermitian disclination states have been studied in topological nonreciprocal scattering networks composed of interconnected unitary circulators, where they emerge in a gap of the anomalous phase [27], to the best of our knowledge, Floquet modes remain unexplored in conservative systems on aperiodic lattices, including systems with disclinations.

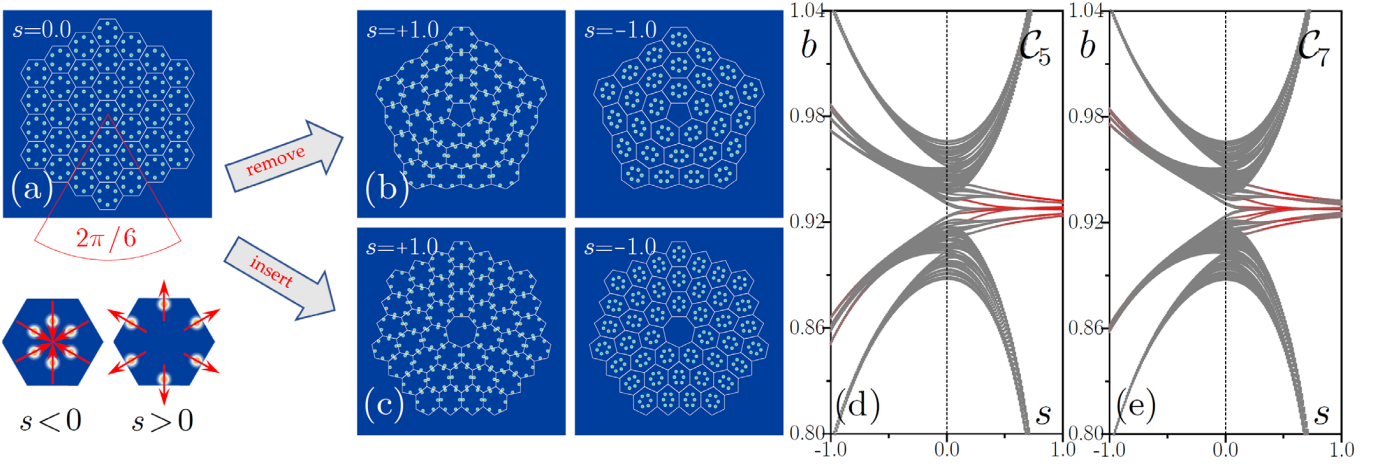

Fig. 1. Original honeycomb waveguide array with $C_6$ discrete rotational symmetry and indicated angular sector with Frank angle $2\pi/6$ that can be removed or added from the array (a). Bottom row illustrates how shift $s$ is introduced into hexagonal unit cells. $C_5$ (b) and $C_7$ (c) waveguide arrays with disclinations corresponding to different shift parameters $s$ that were obtained by removing or adding sector to original $C_6$ structure. Propagation constants of eigenmodes of static $C_5$ (d) and $C_7$ (e) arrays vs shift $s$. The color of dots indicates form-factor $\chi$ of eigenmode (gray – delocalized modes, red – localized modes). Vertical dashed line indicates the point of transition from trivial to topological phase. Here and below $p=6$, $d=3.5$, and $\sigma=0.45$.

In this Letter, we explore a structure that combines the physics of Floquet systems and HOTIs with disclinations, inheriting key characteristics from both classes. We demonstrate that periodic longitudinal modulation of the waveguide positions in the array with disclination can induce Floquet modes localized at the disclination core. These modes are realized in configurations with discrete rotational symmetries $C_5$ and $C_7$, enabling mode profiles unachievable in periodic systems. Direct propagation confirms their Floquet nature, as the modes accurately reproduce their spatial distributions after each modulation period. In the presence of focusing nonlinearity, these Floquet states exhibit robust propagation below a certain power threshold, hinting on the possibility of formation of Floquet solitons on disclinations.

We consider the paraxial propagation of light beam in waveguide array, where waveguides oscillate periodically with distance $z$ in the transverse $(x,y)$ plane. The beam propagation in such medium with focusing nonlinearity is governed by the dimensionless nonlinear Schrödinger equation for field amplitude $\psi(\mathbf{r},z)$:

$$i\frac{\partial \psi}{\partial z} = -\frac{1}{2}\left(\frac{\partial^2 \psi}{\partial x^2} + \frac{\partial^2 \psi}{\partial y^2}\right) - \mathcal{R}(\mathbf{r},z)\psi - |\psi|^2 \psi, \quad (1)$$

where $\mathbf{r}=(x,y)$ are the transverse coordinates, and the function $\mathcal{R}(\mathbf{r},z) = p\sum_{n,m}\mathcal{Q}[x-x_{n,m}(z), y-y_{n,m}(z)]$ describes the refractive-index landscape of the disclination array. It is composed from the identical Gaussian waveguides $\mathcal{Q} = \exp[-(x^2+y^2)/\sigma^2]$ with width $\sigma=0.45$ and depth $p=6$, where the coordinates $x_{n,m}(z), y_{n,m}(z)$ of waveguide centers define the structure with disclination and change periodically with distance $z$.

Disclinations in a *static* ($z$-independent) arrays can be engineered starting from periodic structures, such as finite honeycomb waveguide array that initially possesses $C_6$ discrete rotational symmetry [see Fig. 1(a)]. First, a Kekulé distortion is introduced into all unit cells of the honeycomb array, which initially has a waveguide spacing $d=3.5$, by shifting each waveguide toward the center or toward the periphery of the unit cell by a distance $s$ [see images below Fig. 1(a)]. This deformation modifies the intra- and intercell coupling strengths and is characterized by the ratio $\gamma = l_{\text{intra}}/l_{\text{inter}}$, where $l_{\text{intra}}$ and $l_{\text{inter}}$ are the intracell and intercell spacings, respectively. These quantities are connected with shift $s$ through the following relations $l_{\text{intra}} = 3d\gamma/(1+2\gamma)$ and $l_{\text{inter}} = 3d/(1+2\gamma)$, where $\gamma=(d+s)/(d-2s)$. The unperturbed array corresponds to $s=0$ ($\gamma=1$), while positive $s>0$ and negative $s<0$ shifts lead to topological and trivial phases, respectively. When a sector of the honeycomb array corresponding to the Frank angle of $2\pi/6$ is removed/added [Fig. 1(a)] and positions of remaining waveguides are adjusted to fill/accommodate the removed/added sector, the structure acquires pentagonal ($C_5$) or heptagonal ($C_7$) discrete rotational symmetry with disclination core at the center, as shown in Fig. 1(b) and (c), respectively. Disclination core in the center of the array comprises five waveguides in $C_5$ structure and seven waveguides in $C_7$ structure.

The resulting aperiodic arrays can support topological modes localized at the core for shift $s>0$, and their nontrivial topological properties can be characterized by a fractional spectral charge (see details in [5–7]). Neglecting cubic nonlinearity in Eq. (1), we first obtain linear *static* eigenmodes of the form $\psi=w(\mathbf{r})\exp(ibz)$, where $w$ is a real function describing the modal profile and $b$ is the propagation constant. Propagation constants $b$ of all modes vs shift parameter $s$ are shown in Fig. 1(d) and 1(e) for $C_5$ and $C_7$ arrays, respectively. In addition, the gray-to-red color scale in Fig. 1(d) and 1(e) indicates form-factors $\chi = (\int |w|^4 d^2\mathbf{r})^{1/2}/\int |w|^2 d^2\mathbf{r}$ of all eigenmodes, which quantifies their degree of localization (larger $\chi$ value corresponds to stronger localization of the mode, i.e. gray color corresponds to delocalized states, while red to localized ones). In trivial phase ($s<0$), no states exist within the gap between the bulk bands (gray dots), and all modes of *static* array are extended. In contrast, in topological phase ($s>0$) localized modes emerge within the gap between the bulk bands (red dots). Topological states in this regime may appear not only at the disclination core, but also at outer edges/corners of *static* array (see, e.g., [12]), so part of in-gap red branches in Fig. 1(d), (e) is connected with latter states.

Moving to *Floquet* arrays with disclinations, we now make structure *dynamical* along the longitudinal direction $z$ by varying shift of the waveguides in the unit cell in periodic fashion as $s = r\sin(\omega z)$, where $r$ is the oscillation amplitude, $\omega = 2\pi/Z$ is the oscillation frequency corresponding to the longitudinal period $Z=20$. This modulation causes the center of each waveguide to oscillate radially within its unit cell—that is, along the line connecting the waveguide position and the center of the cell—following the periodic variation of the shift $s(z)$. In this case the system starts at $s=0$, then transits into instantaneous topological phase, returns to $s=0$, and spends the remaining half of the period in the trivial phase. Main finding of this Letter is that although on average $s_{\text{av}} = Z^{-1}\int_0^Z s(z)dz = 0$ and the array may seem to be trivial, in fact it still can support localized Floquet disclination states that appear exclusively due to longitudinal modulation. The eigenmodes of such array are the Floquet states of the form $\psi(\mathbf{r},z) = w(\mathbf{r},z)\exp(ibz)$, where $b \in [-\pi/Z, +\pi/Z]$ is the quasi-propagation constant, and $w(\mathbf{r},z)$ is a $Z$-periodic complex function satisfying $w(\mathbf{r},z) = w(\mathbf{r},z+Z)$. Substituting this ansatz into linear version of Eq. (1), we obtain the linear eigenvalue problem $bw = (1/2)(\partial^2 w/\partial x^2 + \partial^2 w/\partial y^2) + \mathcal{R}w + i\partial w/\partial z$ that can be solved numerically using the propagation and projection method [see Supplementary Material (SM)]. Because longitudinal modulation can lead to the appearance of localized states at the outer corners in the same gap as disclination states, we intentionally removed outer ring of waveguides in dynamical array to simplify its quasi-propagation constant spectrum. Supplementary Visualizations (SV) 1 and 2 illustrate how the profiles of Floquet arrays with disclination evolve over one longitudinal period $Z$ for systems possessing $\mathcal{C}_5$ and $\mathcal{C}_7$ discrete rotational symmetries. Quasi-propagation constants $b$ of Floquet modes of $\mathcal{C}_5$ and $\mathcal{C}_7$ arrays are shown versus amplitude $r$ of waveguide oscillations in Fig. 2(a) and 2(c), respectively, within three longitudinal Brillouin zones to stress periodicity of spectrum along the $b$ axis. Longitudinal modulation opens a gap between the two bands (shown as gray dots), within which disclination states emerge (red dots). This behavior visually resembles that observed in anomalous topological insulators, where topological edge states emerge in proper range of modulation parameters (see, e.g., [16,17,23]), and also the behavior found in modulated Su-Schrieffer-Heeger (SSH) chains supporting $\pi$-modes [24], arising in the points where a band overlaps with its own Floquet replica as a result of the longitudinal modulation. In our case, however, the situation is qualitatively different: the observed Floquet modes arise between two distinct bands of the underlying lattice, rather than from the interaction of a single band with its Floquet replica. Moreover, they occur in a two-dimensional aperiodic geometry with a disclination defect, which introduces additional topological constraints absent in one-dimensional Su-Schrieffer-Heeger systems. Consequently, the gap that hosts these modes is opened directly by the longitudinal modulation acting on the two-dimensional aperiodic structure. For the $\mathcal{C}_5$ configuration [Fig. 2(a)], the spectrum contains two degenerate pairs of disclination modes and one singlet mode, whereas for the $\mathcal{C}_7$ configuration [Fig. 2(c)], three degenerate pairs and one singlet are observed. Thus, the structure of the Floquet spectrum reflects the fundamental constraints imposed by the discrete rotational symmetry of the arrays. The gray-to-red color scale in Fig. 2(a) and 2(c) illustrates the form factor $\chi$ of eigenmodes and it indicates on the enhancement of localization of disclination states with increasing oscillation amplitude $r$.

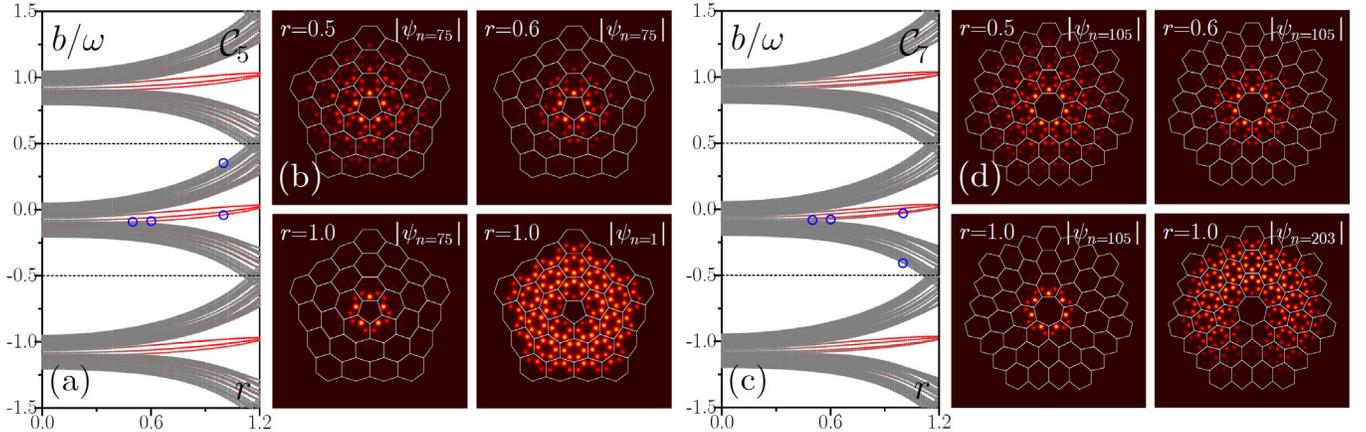

Fig. 2. Quasi-propagation constants of Floquet modes of $\mathcal{C}_5$ (a) and $\mathcal{C}_7$ (c) dynamic waveguide arrays with disclination vs amplitude of waveguide oscillations $r$. Three longitudinal Brillouin zones are shown. Color of dots in Floquet spectrum reflects form-factor $\chi$ of corresponding modes (grey–delocalized modes, red–localized modes). Field modulus distributions $|\psi_n|$ at $z=0Z$ in Floquet modes with different indices $n$ and for different oscillation amplitudes $r$ in $\mathcal{C}_5$ (b) and $\mathcal{C}_7$ (d) arrays corresponding to blue circles in (a) and (c), respectively. Modes are shown within $x,y \in [-40,40]$ window in (b) and $x,y \in [-45,45]$ window in (d). In all cases $Z=20$.

Examples of field modulus distributions $|\psi_n|$ at $z=0Z$ of Floquet states of the structure are presented in Fig. 2(b) for $\mathcal{C}_5$ and in Fig. 2(d) for $\mathcal{C}_7$ arrays for different oscillation amplitudes $r$. In $\mathcal{C}_5$ array we have chosen representative disclination mode with index $n=75$, while in $\mathcal{C}_7$ array the mode $n=105$ is depicted (these modes have equal intensities in waveguides on the disclination core, see also Fig. S1 in SM, which shows all five localized disclination Floquet modes in the $\mathcal{C}_5$ array and all seven localized disclination modes in the $\mathcal{C}_7$ array). As can be seen, increasing the oscillation amplitude $r$ results in stronger localization of disclination modes at the core. Examples of delocalized bulk Floquet modes are shown for $r=1$ too. We note that if the oscillation amplitude becomes too large, i.e. $r > 1.5$, the tails of the waveguides in the array start overlapping, leading to radiation into the bulk of the array.

Floquet disclination modes undergo shape variations during propagation, while remaining exactly $Z$-periodic. This periodicity was verified by simulating propagation of these modes over 100 longitudinal periods in $\mathcal{C}_5$ dynamical array. In Fig. 3(a) we show the evolution of the peak amplitude of mode $n=75$ versus propagation distance $z$. The mode remains stable, showing periodic shape and amplitude variations and minimal radiation due to waveguide curvature, demonstrating its robustness over long-distance propagation. The field distributions at several propagation distances within one modulation period are shown in Fig. 3(c). One can observe that $|\psi|$ exhibits considerable transformation, but the initial profile is restored after each period $Z$. Similar robust behavior was observed for the $\mathcal{C}_7$ lattice (see Fig. S2 in the SM, which confirms the exact $Z$ periodicity of the disclination Floquet modes through the full profile reconstruction on one modulation period and periodic peak amplitude oscillations).

We also investigated the propagation of the Floquet disclination states in the dynamic system with included cubic *nonlinearity*, for different input powers $U = \int |\psi|^2 d^2\mathbf{r}$. The field modulus distributions after $z=100Z$ are shown in Fig. 4(a) for the $\mathcal{C}_5$ structure and input mode $n=75$ and in Fig. 4(b) for the $C_7$ structure and input mode $n=105$. In the nonlinear regime, we observe that below a certain power threshold, the states propagate stably after minor initial adjustments without noticeable radiation. However, above this threshold, which is higher for $C_7$ array than for $C_5$ one, the Floquet disclination state may become unstable, leading to radiation into the bulk and azimuthal instabilities/oscillations of intensity in spots at the disclination core. The observed power threshold may arise because nonlinear self-action shifts the effective quasi-propagation constant of the Floquet disclination mode toward the bulk band. When this nonlinear shift becomes comparable to the gap width, the mode can resonantly couple to extended bulk states, that leads to radiation and may even result in development of azimuthal instabilities. The corresponding evolution dynamics for different powers $U$ are shown in the Supplementary Visualizations (SV) 3-10.

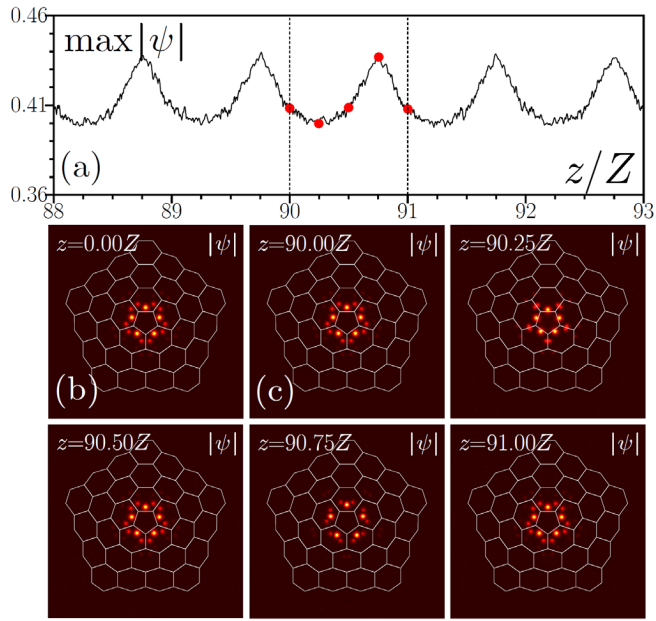

Fig. 3. Evolution dynamics of Floquet state with index $n=75$ in $C_5$ array at $r=1$. Panel (a) shows peak amplitude vs distance $z$, while red dots correspond to intensity distributions shown in (c), the input field distribution is shown in (b).

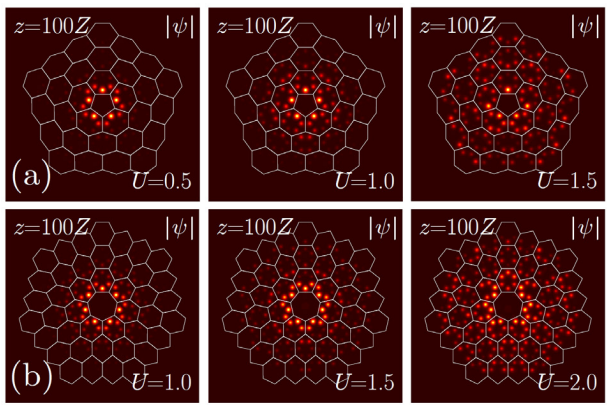

Fig. 4. Output field modulus distributions at $z=100Z$ at different power levels $U$ in $C_5$ array for input mode $n=75$ (a) and in $C_7$ array for input mode $n=105$ (b). In all cases $r=1$. Corresponding evolution dynamics is shown in SV 3-6 for $C_5$ array and SV 7-10 for $C_7$ array.

In summary, we have demonstrated that periodic longitudinal oscillations of waveguides in aperiodic arrays with pentagonal and heptagonal disclination cores can generate Floquet modes localized at the disclination core. These modes reproduce their spatial profiles after each modulation period, confirming their Floquet nature. Nonlinear propagation further reveals that the Floquet disclination states remain localized below a certain power threshold hinting on the possibility of formation of Floquet disclination solitons. Our results offer a new route to topological light confinement arising in aperiodic Floquet systems.

**Funding:** This work was supported by the Russian Science Foundation (grant 24-12-00167) and partially by the research project FFUU-2024-0003 of the Institute of Spectroscopy of the Russian Academy of Sciences. S.K.I. has received funding from the European Union through the Program Fondo Social Europeo Plus 2021-2027(FSE+) of the Valencian Community (Generalitat Valenciana CIAPOS/2023/329).

**Disclosures:** The authors declare no conflicts of interest.

**Data availability.** Data underlying the results presented in this paper may be obtained from the authors upon reasonable request.